\documentclass[conference]{IEEEtran}
\IEEEoverridecommandlockouts
\usepackage{amsmath,amssymb,amsfonts}
\usepackage{graphicx}
\usepackage{textcomp}
\usepackage{enumitem}
\usepackage{balance} 
\usepackage{cite}
\usepackage{xcolor}
\usepackage[numbers]{natbib}
\usepackage[bookmarks=true]{hyperref}

\usepackage{comment}
\usepackage{hhline}
\usepackage{caption}
\usepackage{subcaption}
\usepackage{amsmath}
\usepackage{tikz}
\usetikzlibrary{positioning}
\usepackage[export]{adjustbox}
\usepackage{pgfplots}
\usepackage{pgfplotstable}
\pgfplotsset{compat=1.7}
\usepgfplotslibrary{groupplots}
\usepackage{multirow}
\def\BibTeX{{\rm B\kern-.05em{\sc i\kern-.025em b}\kern-.08em
    T\kern-.1667em\lower.7ex\hbox{E}\kern-.125emX}}
\begin{document}

\newcommand{\newlineauthors}{
  \end{@IEEEauthorhalign}\hfill\mbox{}\par
  \mbox{}\hfill\begin{@IEEEauthorhalign}
}
\makeatother

\title{CORAE: A Tool for Intuitive and Continuous Retrospective Evaluation of Interactions
}

\author{%
  \IEEEauthorblockN{%
    \parbox{\linewidth}{\centering
      Michael J. Sack\IEEEauthorrefmark{1},
      Maria Teresa Parreira\IEEEauthorrefmark{1},
      Jenny Fu\IEEEauthorrefmark{1},
      Asher Lipman\IEEEauthorrefmark{1},
      Hifza Javed\IEEEauthorrefmark{2},
      Nawid Jamali\IEEEauthorrefmark{2}, and
      Malte Jung\IEEEauthorrefmark{1}
    }%
  }%
  \IEEEauthorblockA{%
    \IEEEauthorrefmark{1}Cornell University, 
    \IEEEauthorrefmark{2}Honda Research Institute USA, Inc.
  }%
}

\maketitle
\begin{abstract}
This paper introduces CORAE, a novel web-based open-source tool for \textit{COntinuous Retrospective Affect Evaluation}, designed to capture continuous affect data about interpersonal perceptions in dyadic interactions. Grounded in behavioral ecology perspectives of emotion, this approach replaces valence as the relevant rating dimension with approach and withdrawal, reflecting the degree to which behavior is perceived as increasing or decreasing social distance. We conducted a study to experimentally validate the efficacy of our platform with 24 participants.
The tool's effectiveness was tested in the context of dyadic negotiation, revealing insights about how interpersonal dynamics evolve over time. We find that the continuous affect rating method is consistent with individuals' perception of the overall interaction. This paper contributes to the growing body of research on affective computing and offers a valuable tool for researchers interested in investigating the temporal dynamics of affect and emotion in social interactions.
\end{abstract}

\begin{IEEEkeywords}
Affective computing, interpersonal perception, annotation tool,continuous affect, human-computer interaction
\end{IEEEkeywords}

\IEEEpeerreviewmaketitle

\section{Introduction}

Affect is a dynamic phenomenon. Observable behavior, subjective experience, and physiology all dynamically evolve across time \cite{kuppens2017emotion}. In interactions, affective dynamics co-evolve with those of other interactants \cite{butler2013emotional}, correlating in ways that are yet to be fully explored. Understanding these temporal dynamics requires continuous data streams.

While a growing body of work has addressed strategies and tools to capture affect data over time across physiology \cite{healey2005detecting}, subjective experience \cite{ruef2007continuous, csikszentmihalyi2014validity}, and observable behavior \cite{jung2016thin, coan2007specific}, we lack continuous data about how affective perceptions of others develop dynamically over time.

 Retrospective analysis is an established method to collect continuous data regarding affect \cite{cowie2007feeltrace,cowie2013gtrace,melhart2019pagan,lopes2017ranktrace}. Following this approach, participants are typically video-recorded during an emotion-eliciting event and are later asked to continuously rate how they felt while watching the recording \cite{ruef2007continuous}. This method relies on the phenomenon that individuals often re-experience emotions when reliving a situation \cite{gottman1985valid}. Traditionally, retrospective analysis has focused on collecting continuous data about subjective experience. For example, it was used prominently in couples research by \citet{gottman2000timing}. In this paper, we expand on these works by introducing an approach and a tool that enables researchers to collect continuous affect data about interpersonal perceptions.

Our approach is grounded in a behavioral ecology perspective of emotion, which posits that emotional expressions are primarily social tools used to influence and learn about others \cite{fridlund2014human, crivelli2019inside, van2010emerging}. In line with this perspective, rather than capturing valence when rating interactions, we focus on a dimension of approach and withdrawal, i.e., the degree to which behavior is seen as increasing or decreasing social distance \cite{jung2017affective, caffi1994toward, andersen1996principles}.

 \begin{figure}[ht] 
    \centering
    \includegraphics[width=0.43\textwidth]{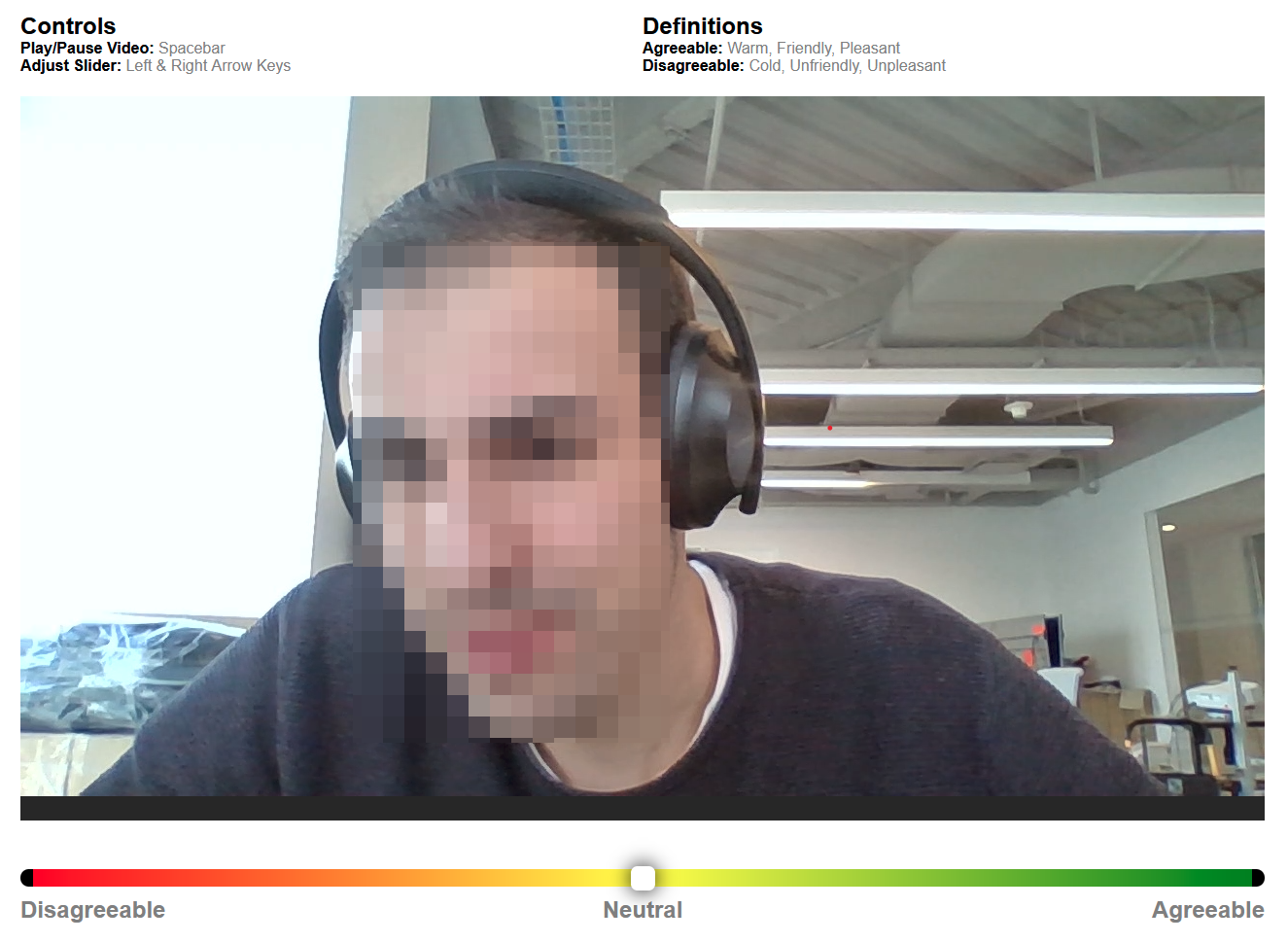}
    \caption{Annotation dashboard for CORAE version 0.15a. After interacting with another individual, participants are asked to retrospectively evaluate how the other person came across by reviewing the recording of the interaction with only the video from the other participant and audio of both.}
    \label{fig:frontend}
\end{figure}

We introduce CORAE, a novel browser-based tool for \textit{\textbf{CO}ntinuous \textbf{R}etrospective \textbf{A}ffect \textbf{E}valuation}. This intuitive tool allows participants to retrospectively rank how another interactant came across immediately following an interaction, thus allowing us to capture interpersonal affective perceptions rather than feelings or affective state inferences. In other words, our system allows us to capture data about how people perceive each other emotionally continuously over time. We make this tool publicly available and test it in a dyadic interaction context, drawing insights about how interpersonal dynamics evolve over time.

Our contributions include:
first, an approach that extends existing affect rating approaches \cite{ruef2007continuous} to capture continuous self-report data about interpersonal affective perceptions.
Second, a novel web-based tool (CORAE) that allows for such data to be captured easily and reliably.
Third, an experimental study where dyads of participants interact and retrospectively rate each other in terms of social distance, which provides evidence that the self-report data captured with our system aligns with interpersonal perceptions people form of others.
 
\section{Related Work}
\label{sec:relatedwork}

\subsection{Measuring affect in interactions}

A growing body of research in affective computing is investigating how to develop and validate reliable measures of affect. Traditionally, affect is represented in two main ways: categorical or dimensional. The first categorizes affective states into discrete emotions (e.g., happiness, sadness) \cite{ekman1992emotions,ekman2003unmasking,lazarus1994emotions}, while the second distinguishes affective states along dimensions such as valence or arousal \cite{mehrabian1996PAD,russel1980circumplex}. Dimensional approaches have been shown to better capture the complexity and nuance of emotional experiences, and can be used to study affective states that do not fit neatly into discrete categories \cite{russel1980circumplex}. To capture these affective states, continuous representations have been popularized \cite{gunes2013trends,mettallinou2013review}, as they allow for an understanding of how humans aggregate affect information across time, unveiling regions of ``emotional saliency'' which may be pivotal to assessing the emotional experience \cite{mettallinou2013review}.

There is also growing interest in the social functions of emotions (e.g., \cite{van2010emerging}). Emotional expressions are increasingly recognized as social tools that people use to influence others and learn about them \cite{fridlund2014human, crivelli2019inside}. From such a behavioral ecology perspective \cite{fridlund2014human, crivelli2019inside}, the expressive meaning of emotions is not pre-determined by specific behavior patterns or morphologies such as facial muscle movements or voice tone patterns, but rather constructed in interaction through a process of "affective grounding" \cite{jung2017affective}. Affective grounding posits that participants of an interaction continuously coordinate on affect to build shared understanding about how behavior should be interpreted affectively and how interaction participants are socially positioned towards each other. This perspective therefore uses ``social distance'' as the relevant rating dimension, evaluating the degree to which social behaviors result in approach and withdrawal \cite{jung2017affective, caffi1994toward, andersen1996principles, lechuga2022comfortability}. Truly understanding how affective grounding is built, however, requires the collection of interpersonal social distance measures \cite{jung2017affective, caffi1994toward, andersen1996principles}, continuously. To the best of our knowledge, this is the first work which addresses retrospective interpersonal affect annotation.

\subsection{Tools for coding affect}
\label{subsec:tools}

Despite significant efforts to develop automated methods for affect recognition \cite{filippini2022deeplearning,wang2019video,zhao2020network}, many studies still rely on human annotators to label affective states in various modalities, such as speech, text, and images. Human annotators provide a level of accuracy and nuance that is difficult to achieve with automated methods, particularly when it comes to complex emotional experiences that may not fit neatly into discrete categories. Furthermore, human annotators can provide valuable insights into the subjective experiences of individuals, enabling researchers to better understand the cognitive and social factors that shape emotional responses.

Nonetheless, manual annotation brings about its own limitations, as several factors hinder validity and reliability. For example, annotators' experience and personal perceptions, the representations chosen, and even the design of the annotation tool \cite{mettallinou2013review} can affect the final labels. Consequently, a wide variety of tools have been developed to address these challenges, including different approaches for continuous affect annotation~\cite{Gottman1985marital,Ruef2007review,yannakakis2015grounding,cowie2007feeltrace, cowie2013gtrace,melhart2019pagan,lopes2017ranktrace, girard2014carma}.

Each of these existing tools affords unique research applications and differs primarily in terms of complexity and accessibility. For example, solutions such as FeelTrace \cite{cowie2007feeltrace}, its successor GTrace \cite{cowie2013gtrace}, and AffectRank \cite{yannakakis2015grounding} each aim to capture multiple dimensions of affect at once. However, these methods can be cognitively demanding \cite{melhart2019pagan}, which is why other approaches, including RankTrace \cite{lopes2017ranktrace}, PAGAN \cite{melhart2019pagan}, and CARMA \cite{girard2014carma} focus more narrowly on a single affective dimension. Further, the mode of annotation for these platforms differ in their use of bounded \cite{cowie2007feeltrace,cowie2013gtrace} and unbounded \cite{lopes2017ranktrace, melhart2019pagan} rating scales. Additionally, these solutions differ in terms of their affordance of in-person and remote data collection; of those mentioned above, all but PAGAN require the collocation of researchers and participants. Conversely, web-based solutions such as PAGAN may theoretically be deployed for remote data collection in addition to in-person contexts.
\section{System Design}
CORAE, related to the Latin word for “heart” and associated with the figurative image of affect, was developed through an iterative design process informed by numerous rounds of pilot testing and user feedback. Broadly, the CORAE platform enables individuals to intuitively evaluate how a person's behavior is interpreted emotionally during interactions. In its current iteration, the platform consists of two components, namely a frontend web interface for participants and a command-line-driven backend to facilitate project management for researchers. Section \ref{subsec:design} details our motivations for various design decisions and situates CORAE in the broader design space. Section \ref{subsec:functionality} details platform functionality implemented as of the experimental validation of CORAE described in Section \ref{sec:study}. Finally, Section \ref{subsec:release} discusses ongoing improvements made to the platform taking insights from researcher and participant feedback. We make CORAE publicly available \footnote{\url{https://corae.org}} as a tool for other researchers.

\subsection{Design}
\label{subsec:design}
Building upon existing tools, we sought to address features and applications under-served by the broader design space and incorporate those most aligned with an intuitive annotation experience. Due to their relative complexity, many existing annotation tools impose barriers to their effective use. Robust as such tools may be, they remain inaccessible to the greater population who lack the necessary training or resources to use them. 

\begin{description}[align=left, leftmargin=0em, labelsep=0.2em, font=\textbf, itemsep=0em,parsep=0.3em]

\item[Layout: ]
We designed CORAE to be intuitive and visually minimal (see \autoref{fig:frontend}) in its presentation. CORAE is deployed in a web browser with the central focus being a video of one's co-interactant, staged for annotation. This is to ensure participants are not distracted by their own image on the screen nor by other visual elements on the platform. Brief instructions above the video player describe keybindings to control the annotation dashboard (\textit{Spacebar} to toggle playback and \textit{Left and Right Arrows} to control the slider), as well as a brief description of the terms used for measuring interpersonal perception (\autoref{fig:instructions}). In the released version of CORAE, the terms used and respective descriptions will be easily customizable, enlarging the use potential for this tool. 

 \begin{figure}[ht] 
    \centering
    \includegraphics[width=0.45\textwidth]{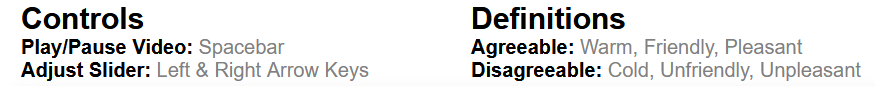}
    \caption{Instruction panel for CORAE version 0.15a. The panel includes a description of available keybindings to control the annotation dashboard (left), as well as the terms used for measuring interpersonal perception (right).}
    \label{fig:instructions}
\end{figure}

 An unobtrusive progress bar is displayed below the video player to inform participants what proportion remains of their evaluation. Finally, below the video player is displayed the annotation slider. A color gradient (from red to green) enables participants to more intuitively understand the meaning of each of the rating terms. The annotation bar is bounded and discretized (a total of 15 points, from $-7$ --- \textbf{Disagreeable} --- to $+7$ --- \textbf{Agreeable}). As mentioned above, this too will be customizable, as other scales may be interesting to explore, namely \textit{Comfortability} \cite{lechuga2022comfortability}. 
 Participants may only change their rating during video playback and are constrained by the platform to do it `continuously'' (i.e., they cannot instantaneously change the rating from \textit{Neutral} ($0$) to \textit{Agreeable} ($7$), but rather adjust to each value in sequence).  

\item[Intuitiveness: ]
By mitigating the need to train annotators on CORAE's use, our platform facilitates continuous retrospective annotation immediately following an interaction. The immediacy of this evaluation allows for stronger salience of affect compared to a delayed approach. Additionally, valuable time that might have otherwise been required to train novice annotators may instead be utilized for more productive ends such as data collection.

\citet{melhart2019pagan} suggest that a lack of intuitiveness can constitute a barrier for researchers such as in the case of tools like FeelTrace \cite{cowie2007feeltrace}, GTrace \cite{cowie2013gtrace}, CARMA \cite{girard2014carma}, and RankTrace \cite{lopes2017ranktrace}. According to the Melhart, the complexity of platform deployment may dissuade researchers from using that platform. 
Melhart's solution, PAGAN~\cite{melhart2019pagan}, is a centralized platform accessible in theory to any interested researcher. In practice, however, the centralized nature of such platforms renders them dependent upon the continued upkeep of a central web server and database. Closed-source solutions further compound this issue by offering little-to-no recourse for researchers who might otherwise desire to deploy their own implementation locally or even to their own web server. Our platform addresses both of these concerns through a well-documented process for third-party deployment and the planned open-source release of our code repository.

\item[Affective Dimension: ]
 The complexity of annotating two or more concurrent affective dimensions imposes a nontrivial cognitive demand upon the annotator \cite{cowie2013gtrace}. This demand, in turn, may diminish the salience of user annotation along both dimensions. Similar to PAGAN \cite{melhart2019pagan} and others, our design aims to capture affect along a single dimension.

\item[Distributed Participation: ]
Whereas existing solutions tend to rely heavily upon the collocation of researchers and participants for in-person data collection (Section \ref{subsec:tools}),  we found this approach to be unnecessarily restrictive and rather limiting to the potential recruitment of more diverse populations. Conversely, remote studies are unable to guarantee consistency for factors such as participant system specifications and environmental distractors. We acknowledge the value afforded by both in-person and remote study formats and sought to develop a tool capable of facilitating either. To this end, CORAE may be deployed locally for in-person sessions as well as remotely for distributed participation.

\end{description}

\subsection{Functionality}
\label{subsec:functionality}
We focused our early development efforts on creating an intuitive and seamless annotation experience for the user. To this end, we withheld several planned features with the intention of their inclusion in a subsequent release of the platform (Section \ref{subsec:release}). When we refer to CORAE and its functionality in this section, it is in reference to build~0.15a unless otherwise noted. The features in this build were those we deemed critical for CORAE's experimental validation and ongoing user testing.

\begin{description}[align=left, leftmargin=0em, labelsep=0.2em, font=\textbf, itemsep=0em,parsep=0.3em]
\item[Annotation Dashboard: ]
 A unique URL is generated for each participant to access their instance of CORAE's annotation dashboard (Fig. \ref{fig:frontend}). Upon accessing their instance, participants are, by default, prompted to enter an identifier which is then logged by the platform and associated with their session. This prompt may be disabled by changing a parameter in the project template. CORAE's annotation dashboard contains three key components: the instruction panel, the video player, and the annotation slider. Platform instructions may be altered in the project template to suit the needs of a study. Dashboard elements responsively scale to account for differences in viewport size and the relative aspect ratio of uploaded media. The annotation slider is labeled at each extreme with the dimension of affect being evaluated.
 
 In terms of interaction, the dashboard affords two primary actions: slider adjustment and playback control. Annotators may indicate their affect rating by adjusting the slider using \textit{Left and Right Arrow} during playback. This affect rating by default is indicated using a continuous 15-point scale but may be changed in the project template to suit any granularity. Although some authors argue in favor of unbounded annotation \cite{lopes2017ranktrace,melhart2019pagan}, we opted to bound ratings as a form of affective grounding across sections. Further, CORAE eliminates the need to hold input controls when adjusting ratings, which we found to cause fatigue among participants. Playback may be paused and resumed using \textit{Spacebar} at any point. Our motivation for constraining interactability to these two actions was to minimize cognitive demand and eliminate any unnecessary sources of distraction during annotation.

\item[Data Logging: ]
Data is logged for a session in two ways: (1) by default, the mode for data logging is set to predetermined intervals of one second, which may be adjusted to any granularity; and (2) to ensure accuracy in the annotation method, CORAE also logs data whenever a change in the rating occurs. Associated data points are the slider position (rating), time code, and video frame (in the format \textit{``SliderNumericalPosition": ``Hours:Minutes:Seconds:VideoFrame"}), which are logged in a JSON file. Given that video is recorded at a rate of 30 frames-per-second, this allows for a resolution of up to $1/30$ s in the annotated data streams.

\subsection{Release Version}
\label{subsec:release}
We intend to continue the development of novel features that improve CORAE's overall utility. One of the most significant additions moving into the next release of our platform is the introduction of a more robust project management system. In the latest build of CORAE, project parameters may be modified, staged, and published using the administration panel both prior to and after deployment. Yet another improvement over our 0.15a release is a server-side pipeline for logging participant data. This data management system aggregates files generated by the platform within a predetermined directory on the web server.

Finally, we note that allowing for the open-source modification of our platform enables researchers and developers to tailor its features to their specific needs beyond those we have anticipated. Further, the potential impact of deprecation is lessened by a public release of CORAE's source code. In the event that we were to discontinue development, CORAE's repository would remain available to access and modify indefinitely.

\end{description}

\section{Experimental validation of CORAE}
\label{sec:study}

We tested a use case for CORAE in the form of an experimental study on interpersonal dynamics during dyadic interactions. We briefly describe the study design below. 

Our study has two aims: a) understanding how participants interact with the platform and b) evaluating the tool's effectiveness in accurately capturing interpersonal affective evaluations over time. This study took place remotely, with two participants interacting digitally while completing a task. The study received IRB approval from Cornell University (IRB0143729).

\subsection{Experimental Procedure}

Participants were recruited through Prolific\footnote{\url{https://www.prolific.co/}}. The study took place fully online. Before scheduling their slot, each participant read and signed a consent form. At the scheduled time for the experiment, both participants received a link to a call on Zencastr\footnote{\url{https://zencastr.com/}}, a video call platform that allows for high-quality recording of each video and audio stream separately. Participants read task instructions, including a description of the discussion topic (Reasons for Poverty task \cite{shek2002reasons}, detailed in Section~\ref{sec:reasons-for-poverty}). After this, participants were recorded while interacting to solve the task. When they reached an agreement, or after 10 minutes of discussion, participants were asked to stop discussing and fill out a survey. This survey collected demographic data, as well as measures of interpersonal affect. In the meantime, the researcher downloaded the data streams, merged the video stream with audio streams from both participants and uploaded them to CORAE. Each participant was then distributed a unique URL which opened an instance of CORAE's annotation platform in their browser. Participants were each presented with a video of their discussion partner were asked to continuously rate \textit{how their partner came across} moment-to-moment. Once finished, participants were instructed to download the annotation file and upload it onto an encrypted database. Finally, after completing an exit survey, participants were compensated for their participation with US\$14, through Prolific. 

\subsection{Reasons for Poverty task}
\label{sec:reasons-for-poverty}

To evaluate our tool, we needed a task that could elicit a broad range of emotions. We used a modified version of the \textit{Reasons for Poverty} task \cite{shek2002reasons}. The task requires participants to rank order a list of ``reasons for poverty'' according to their ``accuracy''. Half of the items follow a reasoning that sees the source of poverty in peoples' situation, i.e., their circumstances, whereas the other half follows a reasoning that sees the source of poverty in peoples' disposition, i.e., their personality. By strategically recruiting participants with opposing beliefs about poverty, we aimed to elicit an emotionally engaging interaction. We used the following instructions:

\textit{You and the other participant must come to an agreement as to a rank of the 5 most relevant causes of poverty in order of the accuracy of each statement. The cause of poverty that is evaluated as being most accurate will be ranked as 1st, and the one that is evaluated as least accurate will be ranked as 5th.}
\textit{
\begin{itemize}
\item Poor people lack the ability to manage money. 
\item Poor people waste their money on inappropriate items. 
\item Poor people do not actively seek to improve their lives. 
\item Poor people lack talents and abilities. 
\item Poor people are exploited by the rich. 
\item The society lacks justice. 
\item Distribution of wealth in society is uneven. 
\item Poor people lack opportunities because they live in poor families. 
\item Poor people live in places where there are not many opportunities. 
\item Poor people have encountered personal misfortunes, which limit their opportunities. 
\item Poor people are discriminated against in society. 
\item Poor people have bad fate. 
\item Poor people lack luck. 
\end{itemize}}

Participants were given a maximum of 10 minutes to discuss, to prevent individuals from getting disengaged when reviewing their discussion on CORAE.

\subsection{Measures}
\subsubsection{User Interaction and Experience}
To evaluate how participants are interacting with the platform, we measured the \textbf{Click Rate} (interval of time between rating data points) and \textbf{Rating Range} (range of values for one retrospective annotation session). To measure user experience, we asked participants for feedback about their interactions with our tool in the exit survey.

\subsubsection{Evaluation Accuracy}
In order to evaluate the effectiveness of our tool in accurately capturing an individual's assessment of the interaction, we compared measures of  \textbf{Interpersonal Agreeableness} and \textbf{Interpersonal Perception (IP)}. The Interpersonal Agreeableness measure was operationalized by asking participants  \textit{``How did the other participant come across?"} on a 7-point Likert scale (from \textit{disagreeable} to \textit{agreeable}) in the post-interaction survey. The Interpersonal Perception measure was operationalized from the continuous interpersonal rating data. First, we increased the resolution of the rating data to a $0.1$s period ($10$Hz), to ease manipulation of the data within and across sessions. Then, to calculate the IP and in line with prior work\cite{jung2016thin, gottman1992marital}, for each participant, we took the cumulative sum of the ratings during the interaction and fitted a linear regression to that data. The Interpersonal Perception (IP) measure is given by the slope of that regression, providing an understanding of how the perception of the other interactant evolved over the interaction. 

\subsubsection{Demographics}
In the post-interaction survey, we collected demographic information (\textbf{age, gender, nationality, race/ethnicity}) and \textbf{personality traits} through the short-version of the Big Five Inventory \cite{rammstedt2007bigfive}. Participants were also asked to rate their \textbf{religiousness} (\textit{not-at-all religious} to \textit{very religious}) and \textbf{political leaning} (\textit{very liberal} to \textit{very conservative}) with 7-point Likert scales. 

\subsection{Participants}

Participants were recruited through Prolific. To elicit disagreement during the interactions, participants were selected according to their political leaning (one conservative- and one liberal-leaning), which they disclose \textit{a priori} on Prolific. Other recruitment criteria were proficiency in English and a computer device with a functioning camera and microphone.
\section{Results}
\label{sec:results}

\subsection{Participants}

A total of 12 interaction sessions (24 participants) were used for this study, with an average interaction duration of $560.01\pm 111.39$ s (total length of interaction data of $13440.22$ s, around 224 minutes). Participants' age ranged from $20-61$ years ($M\pm SD: 39.75 \pm 13.49$). Out of the 24 participants, 13 identified as female, 11 as male. Race/ethnicity was mostly Caucasian/White (17), followed by Asian/Asian American (4), Hispanic/Latino (4) and African/African American/Black (2) (participants could select multiple). Most participants were native speakers of English (21), with 3 proficient users.

\subsection{User Interaction and Experience}
\label{sec:user-interaction-and-experience}

We evaluated how participants interact with the platform during the retrospective annotation sessions. Participants' average \textbf{Click Rate} is $0.42 \pm 0.02 s$. Each interaction had an average of $1116 \pm 222$ rating data points. The rating scale numerically ranges from $-7$ (\textit{disagreeable}) to $+7$ (\textit{agreeable}). For the sessions analyzed, the average \textbf{Rating Range} was $8.08$. The distribution of ratings can be seen in \autoref{fig:ratingdist}.

\begin{figure} 
    \centering
    \includegraphics[width=0.5\textwidth]{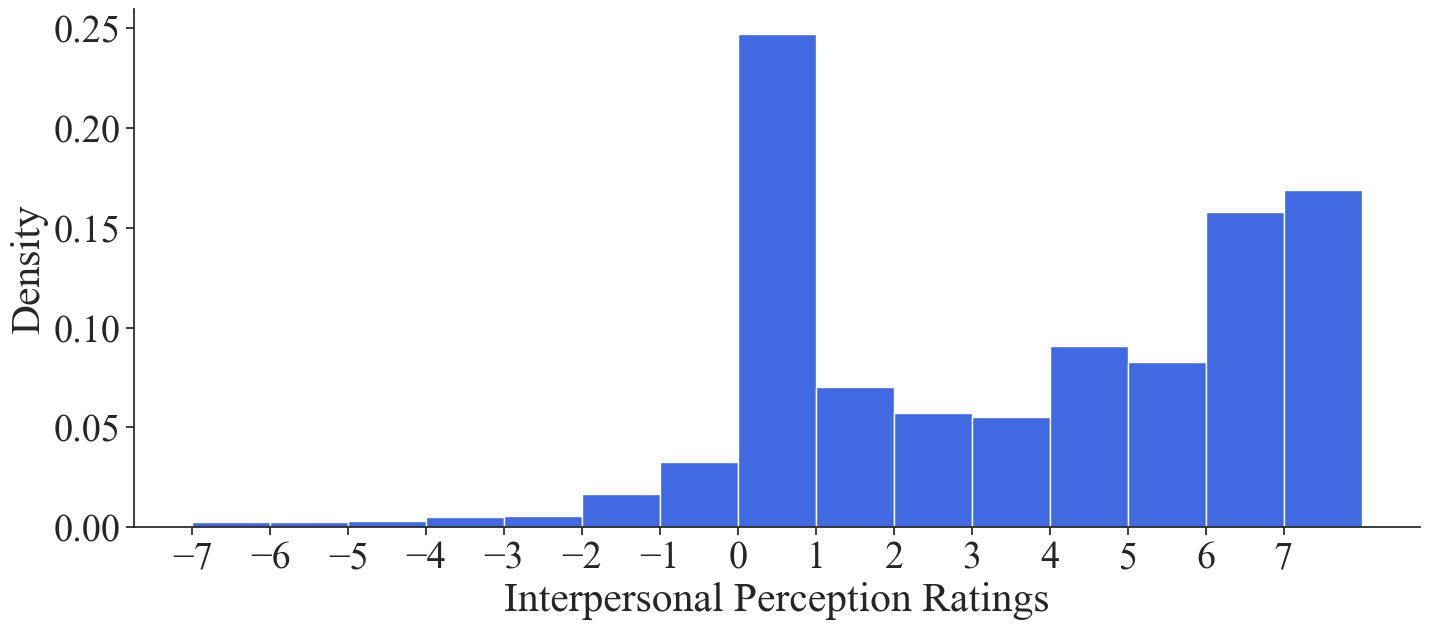}
    \caption{Distribution of rating values in the dataset collected for this study.}
    \label{fig:ratingdist}
\end{figure}

\begin{description}[align=left, leftmargin=0em, labelsep=0.2em, font=\textbf, itemsep=0em,parsep=0.3em]
\item[Feedback from participants:]

Voluntary open-response questions in our exit survey revealed an overall positive annotation experience as reported by users. Given that these questions were open-response and inquired broadly about experiences with the study, respondents, in general, detailed both positive and negative aspects of their interaction during the discussion task rather than with the platform itself. Three (3) participants spoke explicitly to CORAE's \textit{``ease of use''}, and seven (7) reported the platform design to be \textit{``clear'', ``intuitive''}, or some variation thereof. Two (2) participants reported technical difficulties using the playback feature, where the video would not load upon visiting their instance of the annotation dashboard. In both cases, the issue was found to be the result of an unreliable network connection between users and our web server, and the data was still deemed usable for our study. 

\end{description}
\subsection{Evaluation Accuracy}
\label{sec:evaluation-accuracy}

\begin{figure*} [t]
    \centering
    \includegraphics[width=0.85\textwidth]{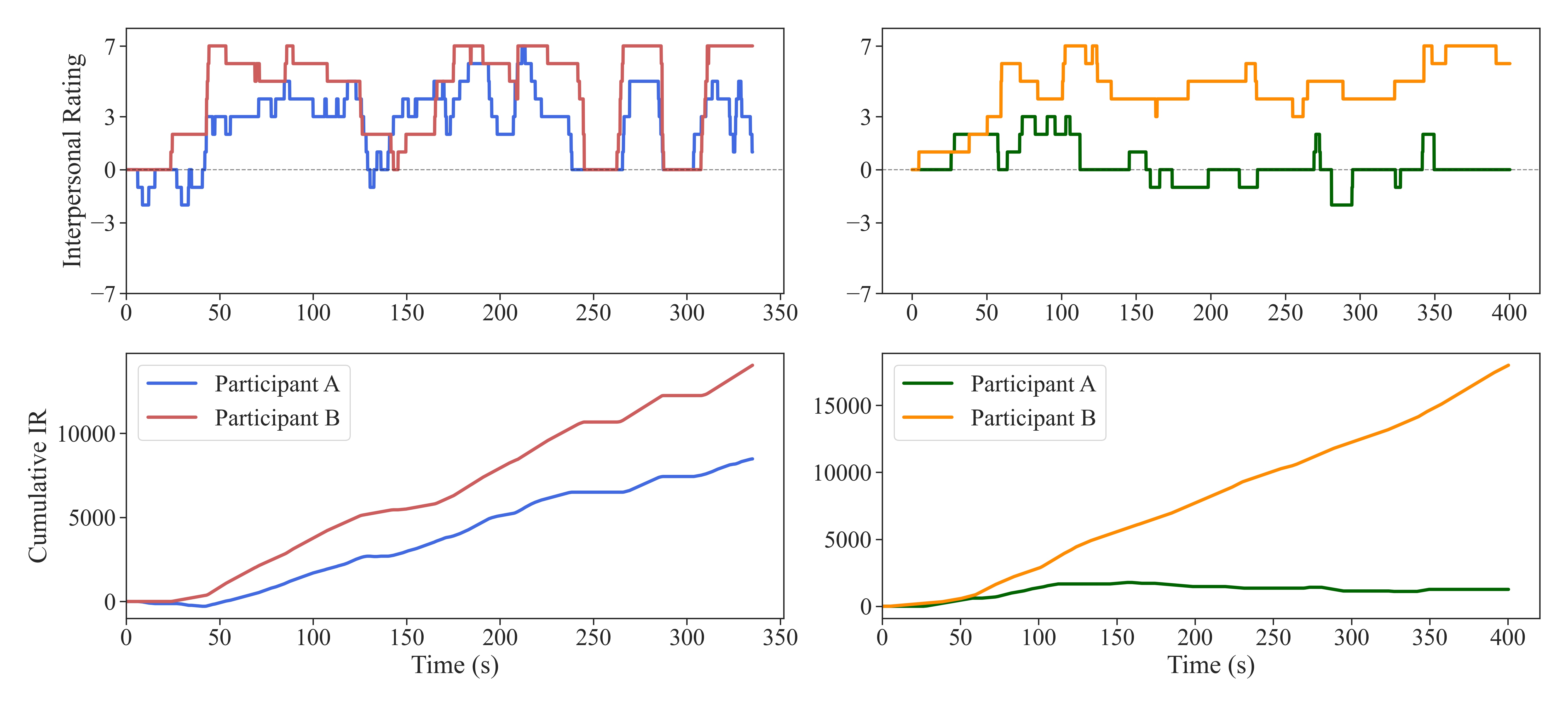}
    \caption{Examples of the dynamics of interpersonal ratings (IR) in two sessions (top) and respective cumulative sum of rated values (bottom). IP is calculated by adjusting a linear regression to the latter curves and extracting the slope of these lines.}
    \label{fig:continuousrate}
\end{figure*}

In \autoref{fig:continuousrate}, we show examples of the retrospective annotation sessions data. Linear regression, used to calculate the \textbf{Interpersonal Perception}, was well-adjusted to the data for all sessions ($R^2:0.88 \pm 0.34$ across the 24 cumulative rating curves). To understand if the retrospective annotation was accurately capturing the perception of the interactants, we calculated the Pearson correlation between the \textbf{Interpersonal Perception} and \textbf{Interpersonal Agreeableness} (assessed by asking the participant to rate in a single scale how the other participant came across). We found a strong correlation between the two measures ($r= 0.72, p<0.001$), indicating good agreement between the continuous rating and the overall assessment of the other participant.

\section{Discussion}
\label{sec:discussion}

This work introduces CORAE, a novel approach to the collection of interpersonal data. CORAE is an intuitive tool for continuous retrospective evaluation of one dimension of the affective content of interactions. Through an evaluation study, we further provide evidence that the tool is both accurate and intuitive/easy to use.
Our results provide insights into individuals' ability to continuously recall affective perceptions of their interaction participant. We also unveil the intricate dynamics of interpersonal perception, with complex data streams that are captured with the use of CORAE as an annotation tool.

CORAE was developed to be intuitive, cutting the need for training sessions and thus increasing the potential for capturing affective states continuously over time. The platform allows for the collection of high-resolution data, with participants changing ratings often and across a broad range of values of interpersonal distance. CORAE thus constitutes a valuable tool that is also easy to customize to other interactions or experimental contexts. We make this tool publicly available and will continue to expand on its functionality.

\autoref{fig:continuousrate} demonstrates how different sessions (and different sets of participants) can entail disparate interpersonal ratings. The plots on the left reflect an interaction where individuals were perceived as mostly agreeable. Interesting synchronization in the ratings can be observed (e.g., the rating drops at around $250 s$ and $300 s$), even though the annotation sessions took place individually. This indicates that a shared understanding of the interaction's affective content may have been established, in line with theories about affective grounding \cite{jung2017affective}. The plots on the right show a session where one participant rated the other much more positively. This becomes particularly apparent when looking at the cumulative IR curves, which reinforces the applicability of the \textbf{Interpersonal Perception} measure as a good single-value indicator of the quality of an interaction as perceived by the individual. In line with this, we found that the IP was strongly correlated with the overall, ``static'' measure of Interpersonal Agreeableness. This shows that CORAE can accurately capture the affective content of an interaction as it relates to social distance. 

Our findings demonstrate the value of using continuous affect rating methods to capture the temporal dynamics of affect and emotion in social interactions. By allowing participants to rate their affective experience both continuously and retrospectively, our tool provides a more fine-grained understanding of the emotional experiences of individuals throughout an interaction. This can help researchers to better understand how affective experiences influence behavior and how behavior, in turn, shapes affective experiences \cite{mettallinou2013review}.

Our approach of replacing valence with approach and withdrawal as the relevant rating dimension reflects a shift towards a more ecologically valid perspective on emotion \cite{fridlund2014human,butler2013emotional,van2010emerging}. This approach acknowledges that emotional experiences are not simply positive or negative but are shaped by the context in which they occur. By focusing on approach and withdrawal, our tool provides a more nuanced understanding of the social dynamics underlying emotional experiences and highlights the importance of considering the social context in which affective experiences occur.

Finally, we believe that CORAE and its subsequent developments have great potential for studying the dynamics of affect and emotion in a range of social contexts beyond dyadic negotiation, such as team collaboration \cite{jung2012team}, or romantic relationships \cite{gottman2000timing, jung2016thin}. Expanding the tool beyond dyadic use could imply a longer annotation process, as each interactant would have to be annotated individually, or pairwise-annotation could occur, where each participant only evaluates one of the interactants; nonetheless, this could provide interesting insights into the differences between individual retrospective evaluation and group interpersonal perception, as the latter requires attention to be split among all the participants. The ability to capture continuous affect data in real-time and retrospectively provides a powerful tool for investigating the complex interplay between affect, behavior, and social context and has the potential to inform the development of more effective interventions and technologies for improving social interactions.

\subsection{Limitations and Future Work}

Despite yielding promising results through our experimental validation of CORAE, we recognize potential limitations to both the platform and broader generalizability of the dataset collected from these sessions. 

\subsubsection{CORAE}

Like any annotation tool, CORAE data might suffer from threats to validity that are common when using human annotators, such as anchoring \cite{seymour2008anchor}, framing \cite{tversky1981framing} and recency \cite{erk2003recency} effects. The use of a retrospective, rather than simultaneous, annotation method can be questioned in terms of validity. Continuous simultaneous rating has been used before \cite{gottman1992marital}, but the cognitive load and meta-affective analysis that needs to take place can hinder the authenticity of the social behaviors demonstrated. The retrospective annotation method addresses these challenges, although the ratings might be affected by future events in an interaction, as individuals may find it hard to abstract from their recalling of how the interaction developed. Nonetheless, our experimental results prove that CORAE ratings effectively reflect interpersonal perceptions of the interactions. Future work may compare simultaneous with retrospective use of the platform. Further, analyses of the affective data can take into account different horizons of the interaction, as insights about how we cognitively model social distance may emerge.

\subsubsection{Experimental Validation}

On the matter of generalizability, constraints relating to participant recruitment skewed our sample heavily in favor of native English speakers from the United States. It would be interesting to see if different cultural backgrounds, which may be attached to different social signaling behaviors \cite{crivelli2019inside}, also correlate with more disagreement in the ratings of interpersonal perceptions. While we kept the analysis of the data focused mostly on the validation of the platform, we see potential in this experiment to shed light on affective grounding mechanisms and implicit and explicit behaviors that signal social distance. For example, disagreement in the ratings may be measured with mean squared error or even inter-rater reliability measures. Future work may also look into personality traits similarities or differences and evaluate if these predict agreement between how participants rated the interaction. Finally, we see potential for the use of machine learning tools to predict interpersonal perception through social behaviors that can be captured through multimodal systems (e.g., audiovisual data).
\section{Conclusion}
\label{sec:conclusion}

In this work, we introduced CORAE, a novel approach to capturing continuous affect data along an approach-withdrawal dimension, reflecting the degree to which behavior is perceived as increasing or decreasing social distance.
Our work contributes to the growing body of research on affective computing and provides a valuable tool for investigating the temporal dynamics of affect and emotion in social interactions. By making our tool publicly available, we hope to promote the adoption of intuitive continuous affect rating methods and facilitate further research into the role of affect in social interactions, namely for machine learning applications. We believe that our tool has the potential to shed light on the complex and nuanced nature of human emotional experiences and inform the development of more effective interventions and technologies for improving social interactions.

\section*{Ethical Impact Statement}
Our study involves human participants, and we have taken several steps to ensure that their privacy and well-being are protected. All participants provided informed consent before participating in the study, and we obtained ethical approval from our institution's ethics board. Participants were duly informed of privacy concerns and our steps to address them. According to Open Science practices, all of the anonymized data used in the current study is made available.
In addition, we recognize that our tool has the potential to be used in a variety of contexts, and that its use could have ethical implications. Models developed from this tool could be used to monitor emotional experiences in workplace interactions, which could potentially lead to negative consequences for employees. We make the tool publicly available with the understanding that researchers who use it will abide by ethical guidelines and take steps to protect the privacy and well-being of their participants. We acknowledge concerns about the use of affective computing technologies and their potential impact. We hope that our research will contribute to a broader conversation about these issues and inform the development of ethical guidelines for the use of affective computing technologies in social interactions.

\section*{Acknowledgment}
This work was supported by Honda Research Institute USA, Inc.. 

\small
\bibliographystyle{abbrvnat}
\balance
\bibliography{bibliography.bib}

\begin{thebibliography}{41}
\providecommand{\natexlab}[1]{#1}
\providecommand{\url}[1]{\texttt{#1}}
\expandafter\ifx\csname urlstyle\endcsname\relax
  \providecommand{\doi}[1]{doi: #1}\else
  \providecommand{\doi}{doi: \begingroup \urlstyle{rm}\Url}\fi

\bibitem[Andersen and Guerrero(1996)]{andersen1996principles}
P.~A. Andersen and L.~K. Guerrero.
\newblock Principles of communication and emotion in social interaction.
\newblock In \emph{Handbook of communication and emotion}, pages 49--96.
  Elsevier, 1996.

\bibitem[Butler and Randall(2013)]{butler2013emotional}
E.~A. Butler and A.~K. Randall.
\newblock Emotional coregulation in close relationships.
\newblock \emph{Emotion Review}, 5\penalty0 (2):\penalty0 202--210, 2013.

\bibitem[Caffi and Janney(1994)]{caffi1994toward}
C.~Caffi and R.~W. Janney.
\newblock Toward a pragmatics of emotive communication.
\newblock \emph{Journal of pragmatics}, 22\penalty0 (3-4):\penalty0 325--373,
  1994.

\bibitem[Coan and Gottman(2007)]{coan2007specific}
J.~A. Coan and J.~M. Gottman.
\newblock The specific affect coding system (spaff).
\newblock \emph{Handbook of emotion elicitation and assessment}, 267:\penalty0
  285, 2007.

\bibitem[Cowie et~al.(2000)Cowie, Douglas-Cowie, Savvidou, McMahon, Sawey, and
  Schr\"oder]{cowie2007feeltrace}
R.~Cowie, E.~Douglas-Cowie, S.~Savvidou, E.~McMahon, M.~Sawey, and
  M.~Schr\"oder.
\newblock 'feeltrace': An instrument for recording perceived emotion in real
  time.
\newblock 01 2000.

\bibitem[Cowie et~al.(2013)Cowie, Sawey, Doherty, Jaimovich, Fyans, and
  Stapleton]{cowie2013gtrace}
R.~Cowie, M.~Sawey, C.~Doherty, J.~Jaimovich, C.~Fyans, and P.~Stapleton.
\newblock Gtrace: General trace program compatible with emotionml.
\newblock In \emph{2013 Humaine Association Conference on Affective Computing
  and Intelligent Interaction}, pages 709--710, 2013.
\newblock \doi{10.1109/ACII.2013.126}.

\bibitem[Crivelli and Fridlund(2019)]{crivelli2019inside}
C.~Crivelli and A.~J. Fridlund.
\newblock Inside-out: From basic emotions theory to the behavioral ecology
  view.
\newblock \emph{Journal of Nonverbal Behavior}, 43\penalty0 (2):\penalty0
  161--194, 2019.

\bibitem[Csikszentmihalyi et~al.(2014)Csikszentmihalyi, Csikszentmihalyi, and
  Larson]{csikszentmihalyi2014validity}
M.~Csikszentmihalyi, M.~Csikszentmihalyi, and R.~Larson.
\newblock Validity and reliability of the experience-sampling method.
\newblock \emph{Flow and the foundations of positive psychology: The collected
  works of Mihaly Csikszentmihalyi}, pages 35--54, 2014.

\bibitem[Ekman(1992)]{ekman1992emotions}
P.~Ekman.
\newblock An argument for basic emotions.
\newblock \emph{Cognition and Emotion}, 6\penalty0 (3-4):\penalty0 169--200,
  1992.
\newblock \doi{10.1080/02699939208411068}.
\newblock URL \url{https://doi.org/10.1080/02699939208411068}.

\bibitem[Ekman and Friesen(2003)]{ekman2003unmasking}
P.~Ekman and W.~Friesen.
\newblock \emph{Unmasking the Face: A Guide to Recognizing Emotions from Facial
  Clues}.
\newblock Number v. 10 in Spectrum book. Malor Books, 2003.
\newblock ISBN 9781883536367.
\newblock URL \url{https://books.google.com/books?id=TukNoJDgMTUC}.

\bibitem[Erk et~al.(2003)Erk, Kiefer, Grothe, Wunderlich, Spitzer, and
  Walter]{erk2003recency}
S.~Erk, M.~Kiefer, J.~Grothe, A.~Wunderlich, M.~Spitzer, and H.~Walter.
\newblock Emotional context modulates subsequent memory effect.
\newblock \emph{NeuroImage}, 18:\penalty0 439--47, 03 2003.
\newblock \doi{10.1016/S1053-8119(02)00015-0}.

\bibitem[Filippini et~al.(2022)Filippini, Di~Crosta, Palumbo, Perpetuini,
  Cardone, Ceccato, Di~Domenico, and Merla]{filippini2022deeplearning}
C.~Filippini, A.~Di~Crosta, R.~Palumbo, D.~Perpetuini, D.~Cardone, I.~Ceccato,
  A.~Di~Domenico, and A.~Merla.
\newblock Automated affective computing based on bio-signals analysis and deep
  learning approach.
\newblock \emph{Sensors}, 22\penalty0 (5), 2022.
\newblock ISSN 1424-8220.
\newblock \doi{10.3390/s22051789}.
\newblock URL \url{https://www.mdpi.com/1424-8220/22/5/1789}.

\bibitem[Fridlund(2014)]{fridlund2014human}
A.~J. Fridlund.
\newblock \emph{Human facial expression: An evolutionary view}.
\newblock Academic press, 2014.

\bibitem[Girard(2014)]{girard2014carma}
J.~M. Girard.
\newblock Carma: Software for continuous affect rating and media annotation.
\newblock \emph{Journal of Open Research Software}, Jul 2014.
\newblock \doi{10.5334/jors.ar}.

\bibitem[Gottman and Levenson(1985{\natexlab{a}})]{Gottman1985marital}
J.~M. Gottman and R.~W. Levenson.
\newblock A valid procedure for obtaining self-report of affect in marital
  interaction.
\newblock \emph{Journal of Consulting and Clinical Psychology}, 53:\penalty0
  151--160, 1985{\natexlab{a}}.

\bibitem[Gottman and Levenson(1985{\natexlab{b}})]{gottman1985valid}
J.~M. Gottman and R.~W. Levenson.
\newblock A valid procedure for obtaining self-report of affect in marital
  interaction.
\newblock \emph{Journal of consulting and clinical psychology}, 53\penalty0
  (2):\penalty0 151, 1985{\natexlab{b}}.

\bibitem[Gottman and Levenson(1992)]{gottman1992marital}
J.~M. Gottman and R.~W. Levenson.
\newblock Marital processes predictive of later dissolution: behavior,
  physiology, and health.
\newblock \emph{Journal of personality and social psychology}, 63\penalty0
  (2):\penalty0 221, 1992.

\bibitem[Gottman and Levenson(2000)]{gottman2000timing}
J.~M. Gottman and R.~W. Levenson.
\newblock The timing of divorce: Predicting when a couple will divorce over a
  14-year period.
\newblock \emph{Journal of Marriage and Family}, 62\penalty0 (3):\penalty0
  737--745, 2000.

\bibitem[Gunes and Schuller(2013)]{gunes2013trends}
H.~Gunes and B.~Schuller.
\newblock Categorical and dimensional affect analysis in continuous input:
  Current trends and future directions.
\newblock \emph{Image and Vision Computing}, 31\penalty0 (2):\penalty0
  120--136, 2013.
\newblock ISSN 0262-8856.
\newblock \doi{https://doi.org/10.1016/j.imavis.2012.06.016}.
\newblock URL
  \url{https://www.sciencedirect.com/science/article/pii/S0262885612001084}.
\newblock Affect Analysis In Continuous Input.

\bibitem[Healey and Picard(2005)]{healey2005detecting}
J.~A. Healey and R.~W. Picard.
\newblock Detecting stress during real-world driving tasks using physiological
  sensors.
\newblock \emph{IEEE Transactions on intelligent transportation systems},
  6\penalty0 (2):\penalty0 156--166, 2005.

\bibitem[Jung et~al.(2012)Jung, Chong, and Leifer]{jung2012team}
M.~Jung, J.~Chong, and L.~Leifer.
\newblock Group hedonic balance and pair programming performance: Affective
  interaction dynamics as indicators of performance.
\newblock In \emph{Proceedings of the SIGCHI Conference on Human Factors in
  Computing Systems}, CHI '12, page 829–838, New York, NY, USA, 2012.
  Association for Computing Machinery.
\newblock ISBN 9781450310154.
\newblock \doi{10.1145/2207676.2208523}.
\newblock URL \url{https://doi.org/10.1145/2207676.2208523}.

\bibitem[Jung(2016)]{jung2016thin}
M.~F. Jung.
\newblock Coupling interactions and performance: Predicting team performance
  from thin slices of conflict.
\newblock \emph{ACM Trans. Comput.-Hum. Interact.}, 23\penalty0 (3), jun 2016.
\newblock ISSN 1073-0516.
\newblock \doi{10.1145/2753767}.
\newblock URL \url{https://doi.org/10.1145/2753767}.

\bibitem[Jung(2017)]{jung2017affective}
M.~F. Jung.
\newblock Affective grounding in human-robot interaction.
\newblock In \emph{Proceedings of the 2017 ACM/IEEE International Conference on
  Human-Robot Interaction}, pages 263--273, 2017.

\bibitem[Kuppens and Verduyn(2017)]{kuppens2017emotion}
P.~Kuppens and P.~Verduyn.
\newblock Emotion dynamics.
\newblock \emph{Current Opinion in Psychology}, 17:\penalty0 22--26, 2017.

\bibitem[Lazarus and Lazarus(1994)]{lazarus1994emotions}
R.~S. Lazarus and B.~N. Lazarus.
\newblock \emph{Passion and Reason: Making Sense of Our Emotions}.
\newblock Oxford University Press USA, 1994.

\bibitem[Lechuga~Redondo et~al.(2022)Lechuga~Redondo, Niewiadomski, Francesco,
  and Sciutti]{lechuga2022comfortability}
M.~E. Lechuga~Redondo, R.~Niewiadomski, R.~Francesco, and A.~Sciutti.
\newblock Comfortability recognition from visual non-verbal cues.
\newblock In \emph{Proceedings of the 2022 International Conference on
  Multimodal Interaction}, ICMI '22, page 207–216, New York, NY, USA, 2022.
  Association for Computing Machinery.
\newblock ISBN 9781450393904.
\newblock \doi{10.1145/3536221.3556631}.
\newblock URL \url{https://doi.org/10.1145/3536221.3556631}.

\bibitem[Lopes et~al.(2017)Lopes, Yannakakis, and Liapis]{lopes2017ranktrace}
P.~Lopes, G.~N. Yannakakis, and A.~Liapis.
\newblock Ranktrace: Relative and unbounded affect annotation.
\newblock In \emph{2017 Seventh International Conference on Affective Computing
  and Intelligent Interaction (ACII)}, pages 158--163, 2017.
\newblock \doi{10.1109/ACII.2017.8273594}.

\bibitem[Mehrabian(1996)]{mehrabian1996PAD}
A.~Mehrabian.
\newblock Pleasure-arousal-dominance: A general framework for describing and
  measuring individual differences in temperament.
\newblock \emph{Current Psychology}, 14:\penalty0 261--292, 1996.

\bibitem[Melhart et~al.(2019)Melhart, Liapis, and Yannakakis]{melhart2019pagan}
D.~Melhart, A.~Liapis, and G.~N. Yannakakis.
\newblock Pagan: Video affect annotation made easy.
\newblock pages 130--136. IEEE, 9 2019.
\newblock ISBN 978-1-7281-3888-6.
\newblock \doi{10.1109/ACII.2019.8925434}.
\newblock URL \url{https://ieeexplore.ieee.org/document/8925434/}.

\bibitem[Metallinou and Narayanan(2013)]{mettallinou2013review}
A.~Metallinou and S.~Narayanan.
\newblock Annotation and processing of continuous emotional attributes:
  Challenges and opportunities.
\newblock In \emph{2013 10th IEEE International Conference and Workshops on
  Automatic Face and Gesture Recognition (FG)}, pages 1--8, 2013.
\newblock \doi{10.1109/FG.2013.6553804}.

\bibitem[Rammstedt and John(2007)]{rammstedt2007bigfive}
B.~Rammstedt and O.~P. John.
\newblock Measuring personality in one minute or less: A 10-item short version
  of the big five inventory in english and german.
\newblock \emph{Journal of Research in Personality}, 41\penalty0 (1):\penalty0
  203--212, 2007.
\newblock ISSN 0092-6566.
\newblock \doi{https://doi.org/10.1016/j.jrp.2006.02.001}.
\newblock URL
  \url{https://www.sciencedirect.com/science/article/pii/S0092656606000195}.

\bibitem[Ruef and Levenson(2007{\natexlab{a}})]{Ruef2007review}
A.~Ruef and R.~Levenson.
\newblock Continuous measurement of emotion: The affect rating dial.
\newblock In J.~Coan and J.~Allen, editors, \emph{Handbook of emotion
  elicitation and assessment}. Oxford University Press, New York, Ny,
  2007{\natexlab{a}}.

\bibitem[Ruef and Levenson(2007{\natexlab{b}})]{ruef2007continuous}
A.~M. Ruef and R.~W. Levenson.
\newblock Continuous measurement of emotion.
\newblock \emph{Handbook of emotion elicitation and assessment}, pages
  286--297, 2007{\natexlab{b}}.

\bibitem[Russell(1980)]{russel1980circumplex}
J.~Russell.
\newblock A circumplex model of affect.
\newblock \emph{Journal of Personality and Social Psychology}, 39:\penalty0
  1161--1178, 12 1980.
\newblock \doi{10.1037/h0077714}.

\bibitem[Seymour and McClure(2008)]{seymour2008anchor}
B.~Seymour and S.~M. McClure.
\newblock Anchors, scales and the relative coding of value in the brain.
\newblock \emph{Current Opinion in Neurobiology}, 18\penalty0 (2):\penalty0
  173--178, 2008.
\newblock ISSN 0959-4388.
\newblock \doi{https://doi.org/10.1016/j.conb.2008.07.010}.
\newblock URL
  \url{https://www.sciencedirect.com/science/article/pii/S0959438808000676}.
\newblock Cognitive neuroscience.

\bibitem[Shek(2002)]{shek2002reasons}
D.~T.~L. Shek.
\newblock Chinese adolescent' explanations of poverty: the perceived causes of
  poverty scale.
\newblock \emph{Adolescence}, 37:\penalty0 789--804, 12 2002.
\newblock ISSN 00018449.

\bibitem[Tversky and Kahneman(1981)]{tversky1981framing}
A.~Tversky and D.~Kahneman.
\newblock The framing of decisions and the psychology of choice.
\newblock \emph{Science}, 211\penalty0 (4481):\penalty0 453--458, 1981.
\newblock \doi{10.1126/science.7455683}.
\newblock URL \url{https://www.science.org/doi/abs/10.1126/science.7455683}.

\bibitem[Van~Kleef(2010)]{van2010emerging}
G.~A. Van~Kleef.
\newblock The emerging view of emotion as social information.
\newblock \emph{Social and Personality Psychology Compass}, 4\penalty0
  (5):\penalty0 331--343, 2010.

\bibitem[Wang et~al.(2019)Wang, Chen, and Ji]{wang2019video}
S.~Wang, S.~Chen, and Q.~Ji.
\newblock Content-based video emotion tagging augmented by users’ multiple
  physiological responses.
\newblock \emph{IEEE Transactions on Affective Computing}, 10\penalty0
  (2):\penalty0 155--166, 2019.
\newblock \doi{10.1109/TAFFC.2017.2702749}.

\bibitem[Yannakakis and Martínez(2015)]{yannakakis2015grounding}
G.~N. Yannakakis and H.~P. Martínez.
\newblock Grounding truth via ordinal annotation.
\newblock In \emph{2015 International Conference on Affective Computing and
  Intelligent Interaction (ACII)}, pages 574--580, 2015.
\newblock \doi{10.1109/ACIwI.2015.7344627}.

\bibitem[Zhao et~al.(2020)Zhao, Ma, Gu, Yang, Xing, Xu, Hu, Chai, and
  Keutzer]{zhao2020network}
S.~Zhao, Y.~Ma, Y.~Gu, J.~Yang, T.~Xing, P.~Xu, R.~Hu, H.~Chai, and K.~Keutzer.
\newblock An end-to-end visual-audio attention network for emotion recognition
  in user-generated videos.
\newblock \emph{Proceedings of the AAAI Conference on Artificial Intelligence},
  34\penalty0 (01):\penalty0 303--311, Apr. 2020.
\newblock \doi{10.1609/aaai.v34i01.5364}.
\newblock URL \url{https://ojs.aaai.org/index.php/AAAI/article/view/5364}.

\end{thebibliography}

\end{document}


\title{Supplementary Material for \\ ``CORAE: A Tool for Intuitive and Continuous Retrospective Evaluation of Interactions''
}

\author{\IEEEauthorblockN{Anonymous Author(s)} \\
\IEEEauthorblockA{\textit{Anonymous Institution(s)}}}
\maketitle

Below, we leave data collected from the annotations using CORAE, for all the sessions included in this study. Additionally, we show the cumulative rating of each session and the corresponding linear regression fit.

\begin{figure*} [h]
    \centering
    \includegraphics[width=0.9\textwidth]{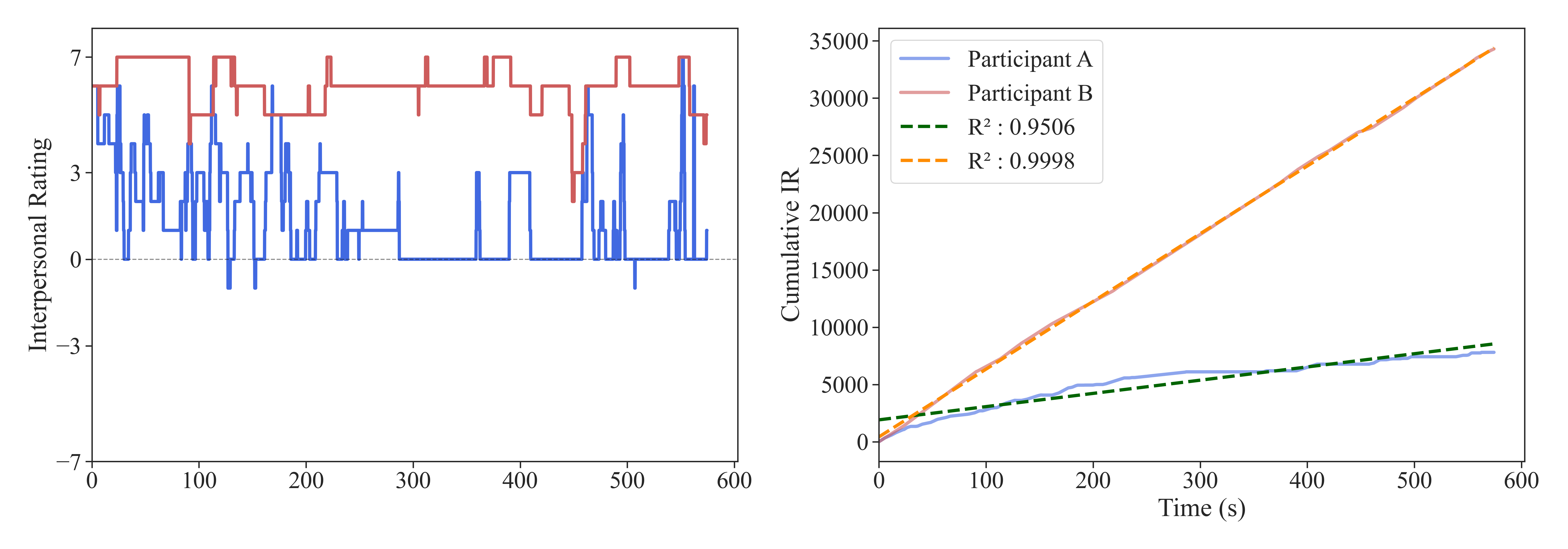}
    \caption{Dynamics of interpersonal ratings (IR) in one session (left) and respective cumulative sum of rated values (right). Interpersonal Perception is calculated by adjusting a linear regression to the latter curves, and extracting the slope of these lines (right, overlaid dashed curve).}
    \label{fig:continuousrate}
\end{figure*}

\begin{figure*} [h]
    \centering
    \includegraphics[width=0.9\textwidth]{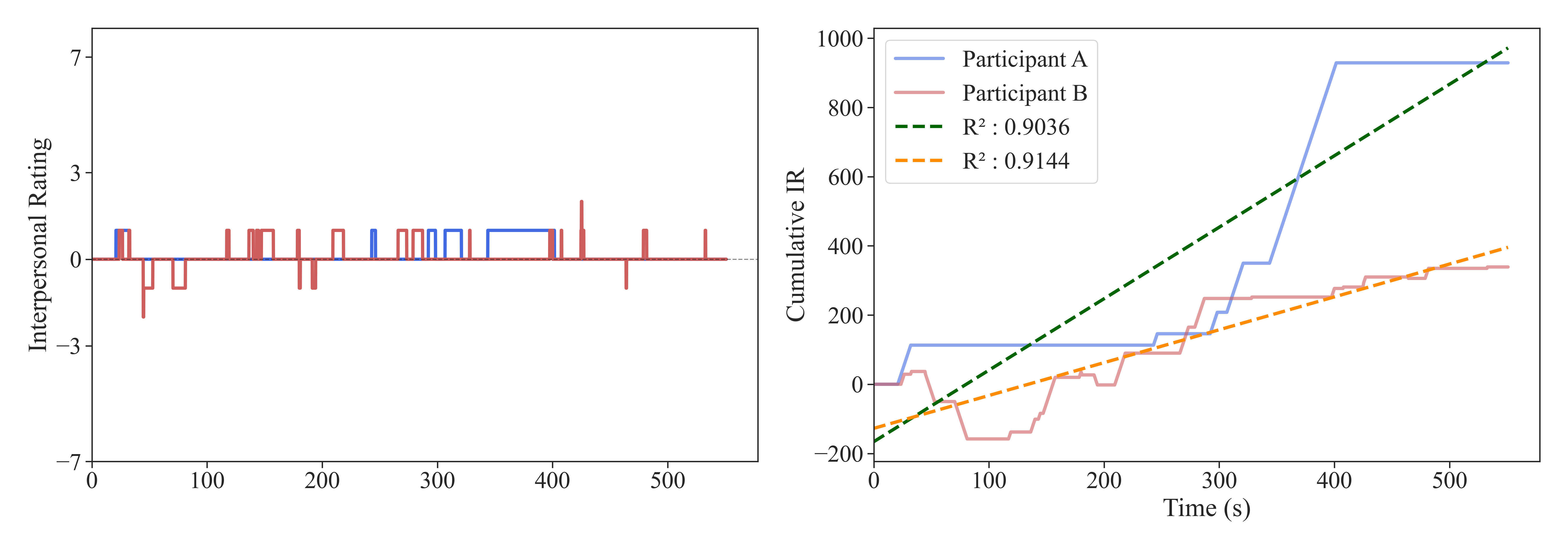}
    \caption{Dynamics of interpersonal ratings (IR) in one session (left) and respective cumulative sum of rated values (right). Interpersonal Perception is calculated by adjusting a linear regression to the latter curves, and extracting the slope of these lines (right, overlaid dashed curve).}
    \label{fig:continuousrate}
\end{figure*}

\begin{figure*} [h]
    \centering
    \includegraphics[width=0.9\textwidth]{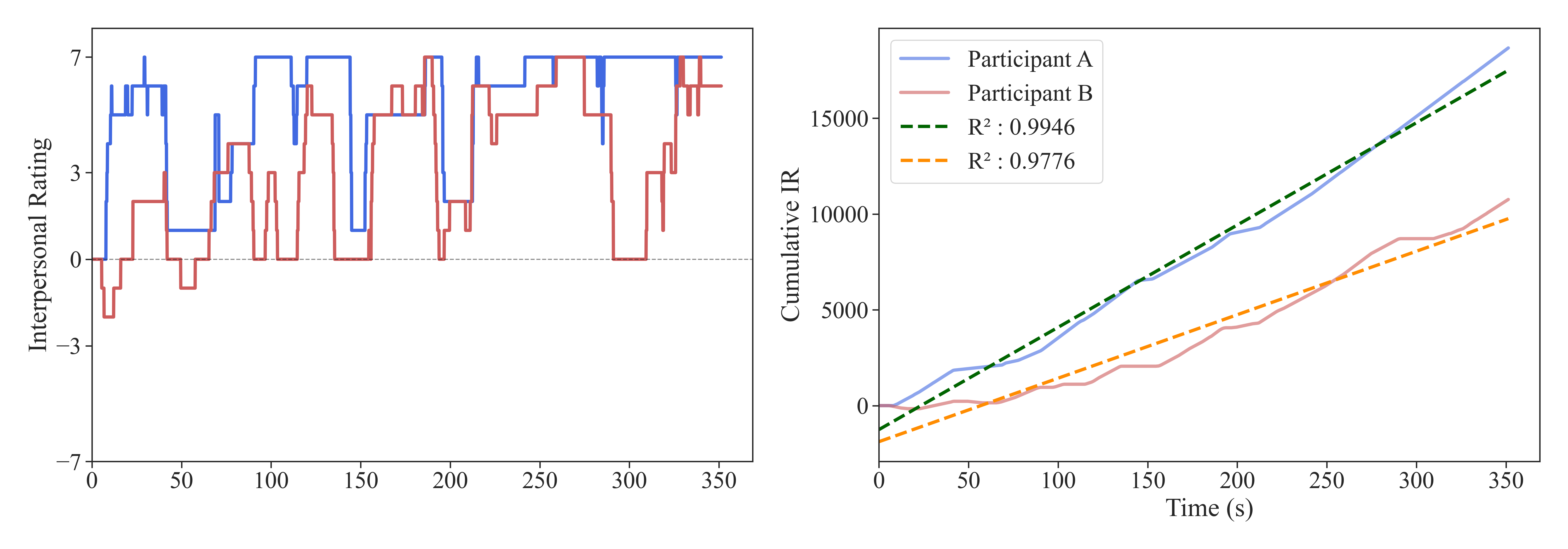}
    \caption{Dynamics of interpersonal ratings (IR) in one session (left) and respective cumulative sum of rated values (right). Interpersonal Perception is calculated by adjusting a linear regression to the latter curves, and extracting the slope of these lines (right, overlaid dashed curve).}
    \label{fig:continuousrate}
\end{figure*}

\begin{figure*} [h]
    \centering
    \includegraphics[width=0.9\textwidth]{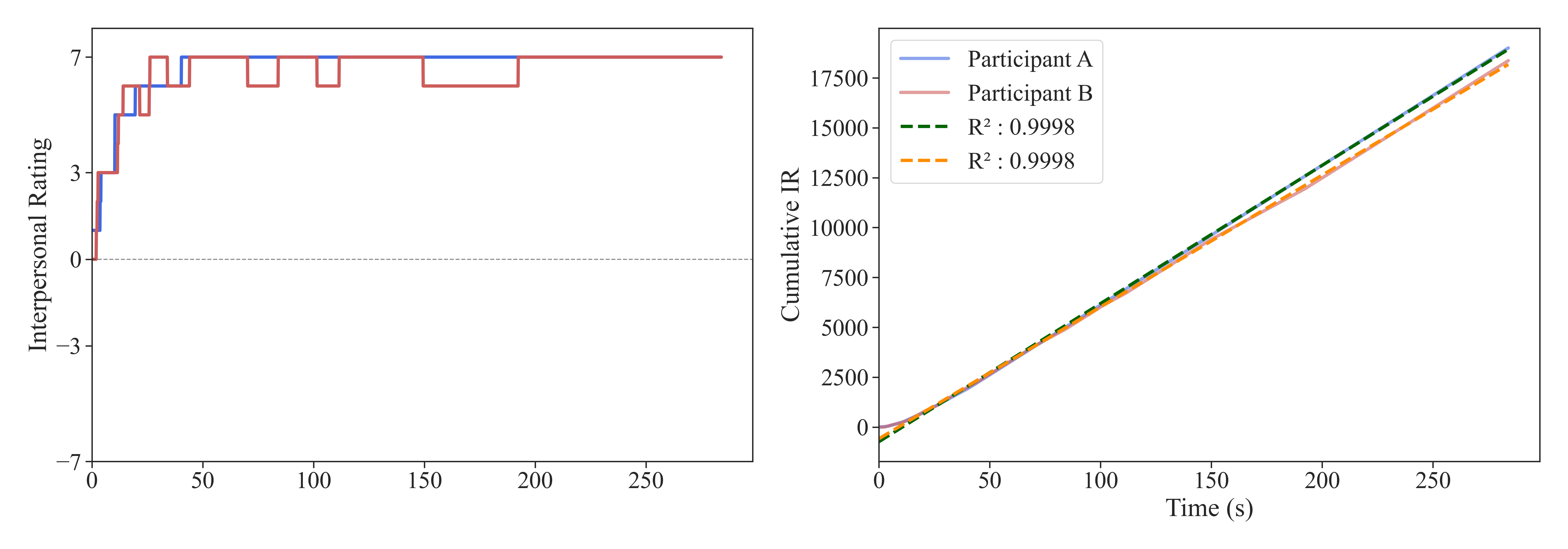}
    \caption{Dynamics of interpersonal ratings (IR) in one session (left) and respective cumulative sum of rated values (right). Interpersonal Perception is calculated by adjusting a linear regression to the latter curves, and extracting the slope of these lines (right, overlaid dashed curve).}
    \label{fig:continuousrate}
\end{figure*}

\begin{figure*} [h]
    \centering
    \includegraphics[width=0.9\textwidth]{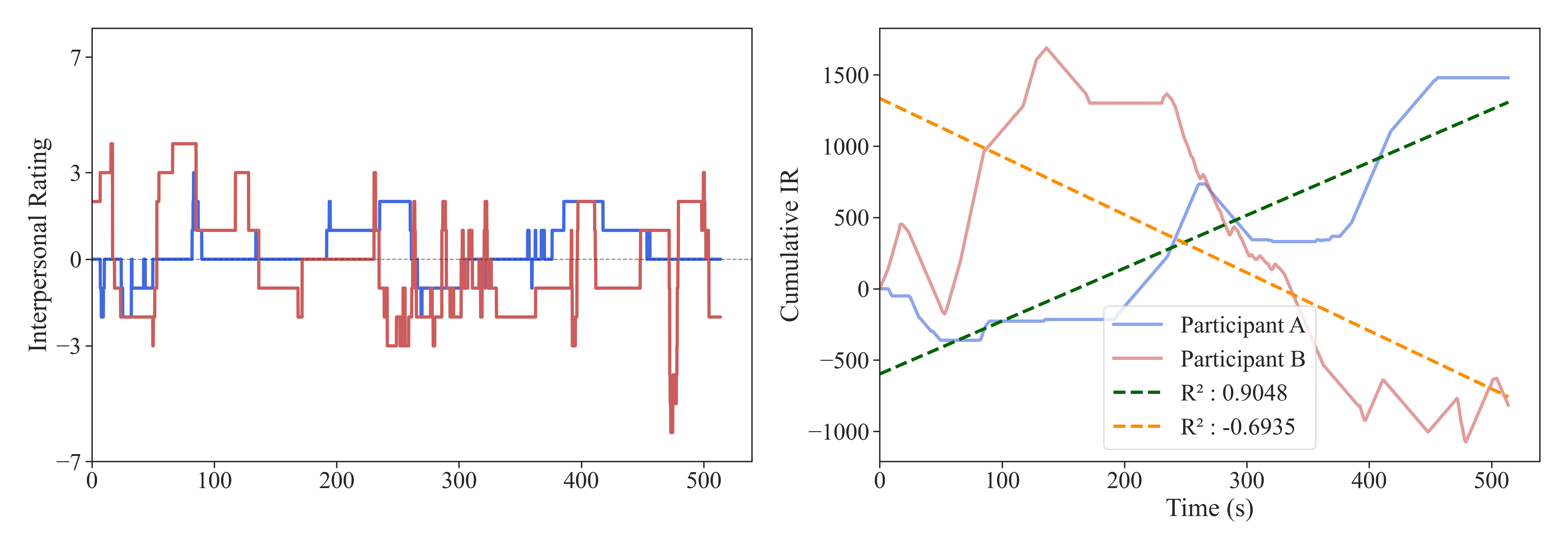}
    \caption{Dynamics of interpersonal ratings (IR) in one session (left) and respective cumulative sum of rated values (right). Interpersonal Perception is calculated by adjusting a linear regression to the latter curves, and extracting the slope of these lines (right, overlaid dashed curve).}
    \label{fig:continuousrate}
\end{figure*}

\begin{figure*} [h]
    \centering
    \includegraphics[width=0.9\textwidth]{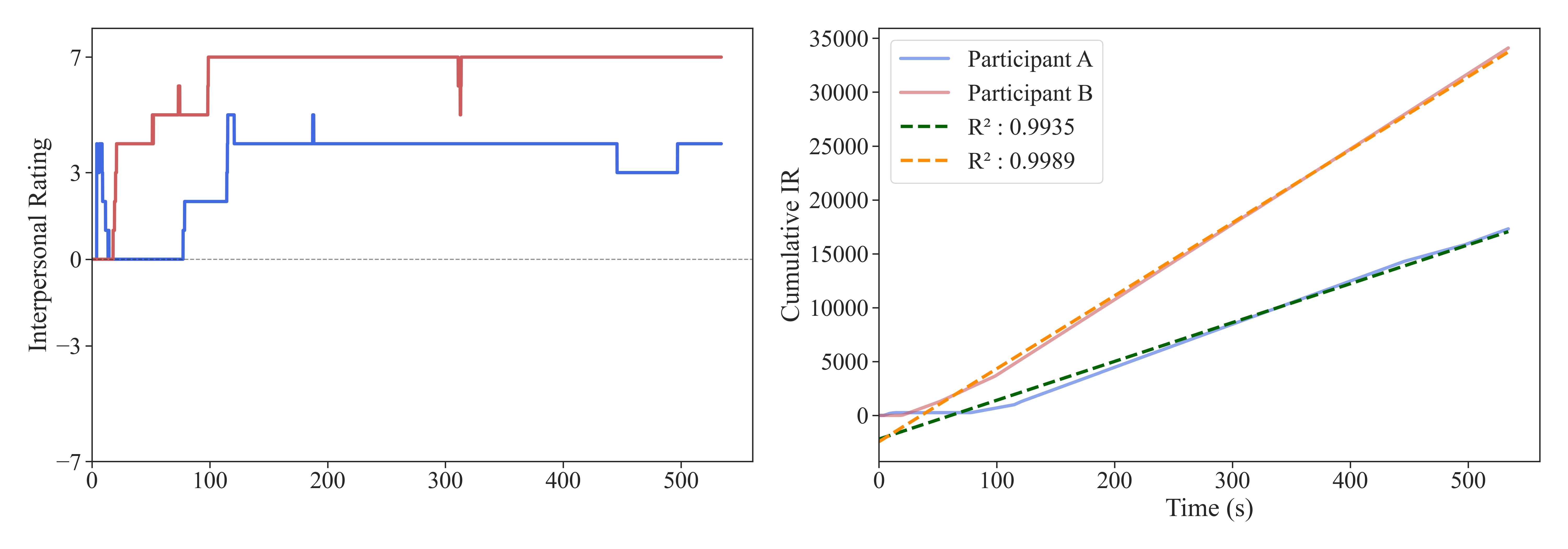}
    \caption{Dynamics of interpersonal ratings (IR) in one session (left) and respective cumulative sum of rated values (right). Interpersonal Perception is calculated by adjusting a linear regression to the latter curves, and extracting the slope of these lines (right, overlaid dashed curve).}
    \label{fig:continuousrate}
\end{figure*}

\begin{figure*} [h]
    \centering
    \includegraphics[width=0.9\textwidth]{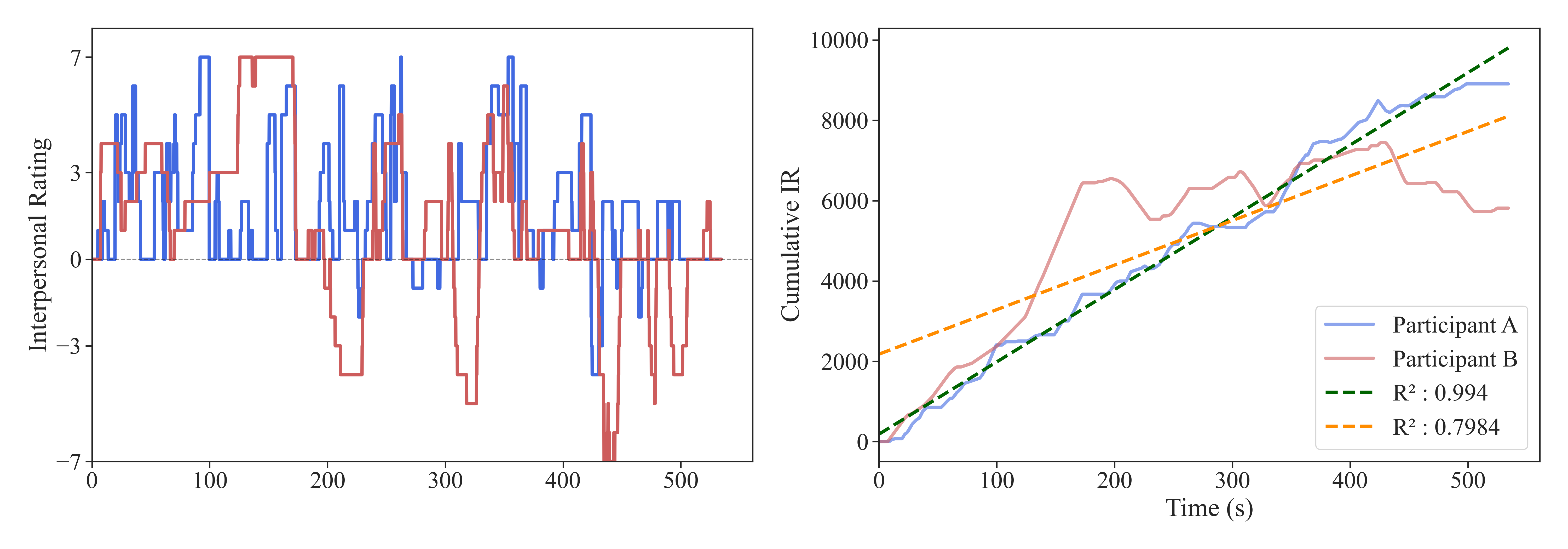}
    \caption{Dynamics of interpersonal ratings (IR) in one session (left) and respective cumulative sum of rated values (right). Interpersonal Perception is calculated by adjusting a linear regression to the latter curves, and extracting the slope of these lines (right, overlaid dashed curve).}
    \label{fig:continuousrate}
\end{figure*}

\begin{figure*} [h]
    \centering
    \includegraphics[width=0.9\textwidth]{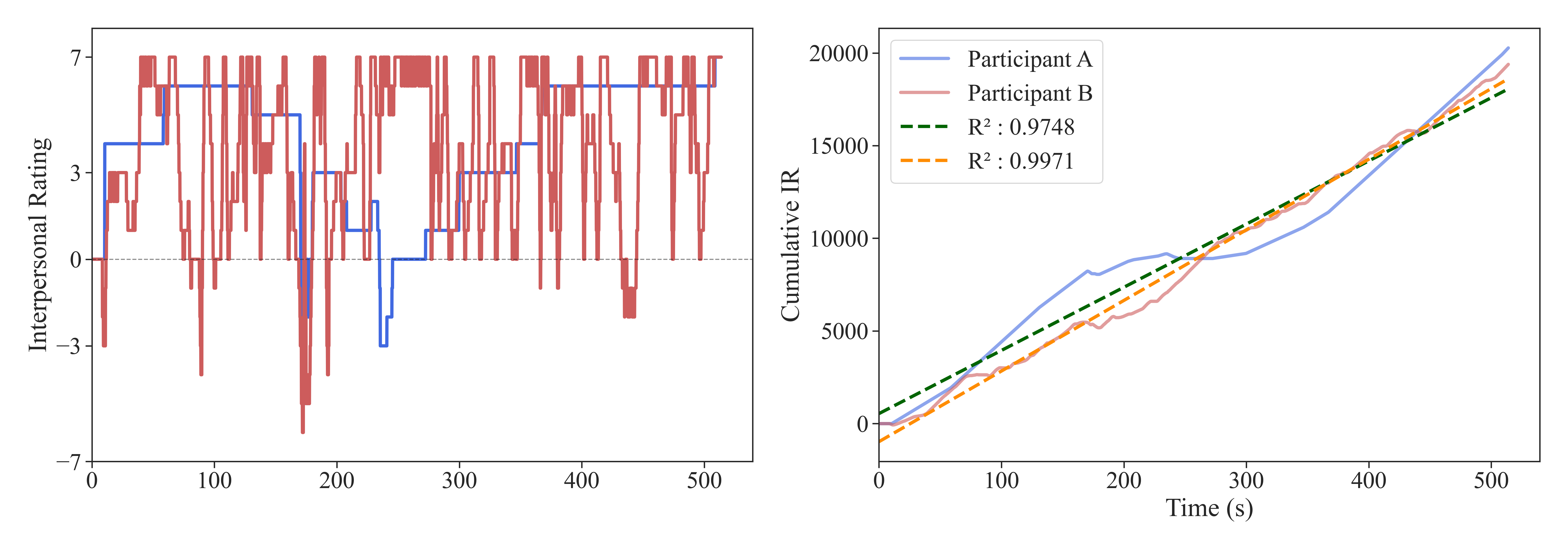}
    \caption{Dynamics of interpersonal ratings (IR) in one session (left) and respective cumulative sum of rated values (right). Interpersonal Perception is calculated by adjusting a linear regression to the latter curves, and extracting the slope of these lines (right, overlaid dashed curve).}
    \label{fig:continuousrate}
\end{figure*}

\begin{figure*} [h]
    \centering
    \includegraphics[width=0.9\textwidth]{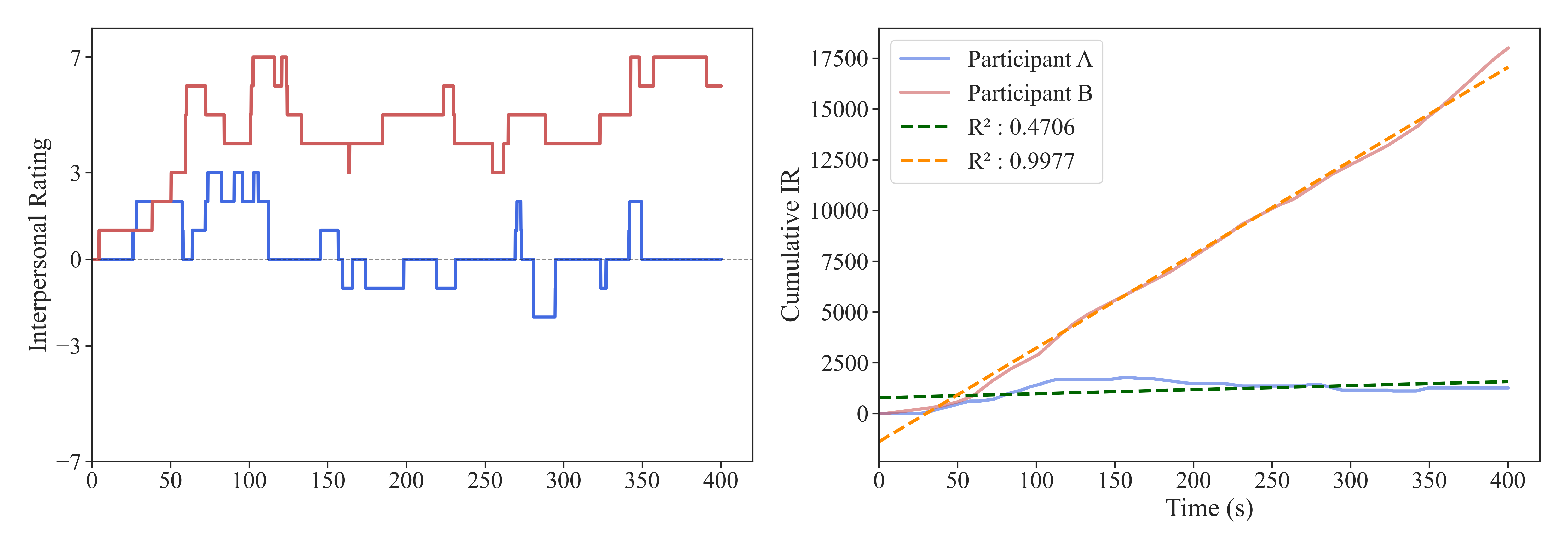}
    \caption{Dynamics of interpersonal ratings (IR) in one session (left) and respective cumulative sum of rated values (right). Interpersonal Perception is calculated by adjusting a linear regression to the latter curves, and extracting the slope of these lines (right, overlaid dashed curve).}
    \label{fig:continuousrate}
\end{figure*}

\begin{figure*} [h]
    \centering
    \includegraphics[width=0.9\textwidth]{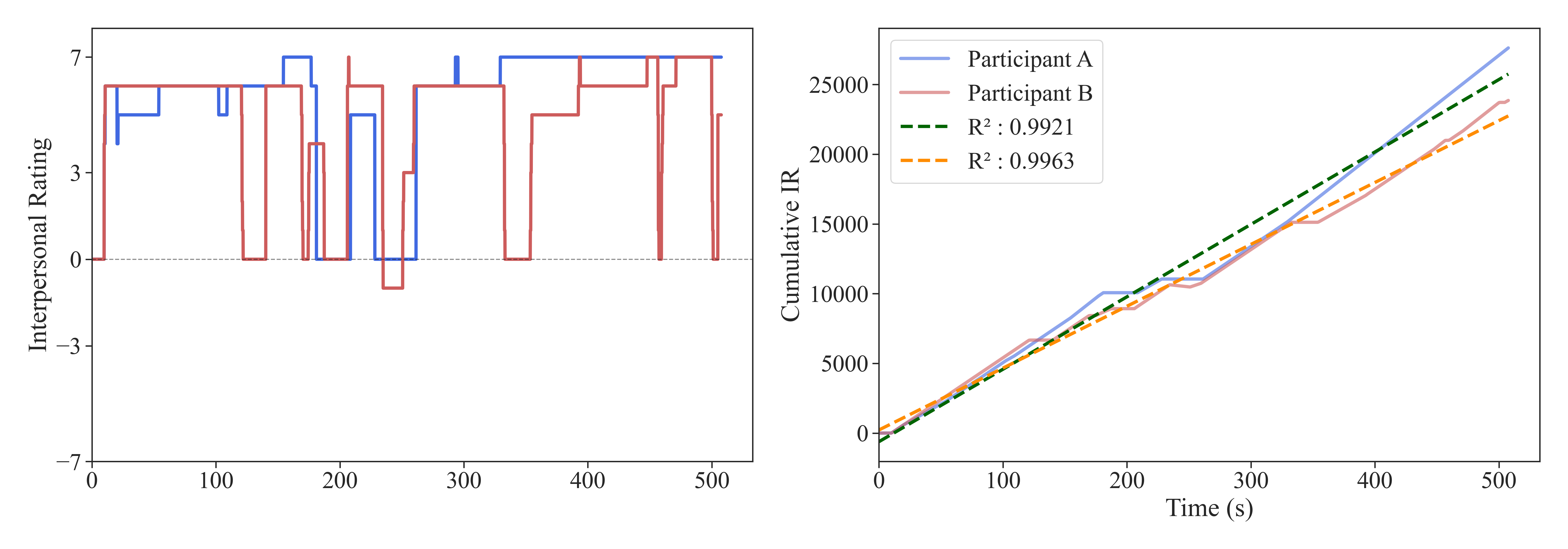}
    \caption{Dynamics of interpersonal ratings (IR) in one session (left) and respective cumulative sum of rated values (right). Interpersonal Perception is calculated by adjusting a linear regression to the latter curves, and extracting the slope of these lines (right, overlaid dashed curve).}
    \label{fig:continuousrate}
\end{figure*}

\begin{figure*} [h]
    \centering
    \includegraphics[width=0.9\textwidth]{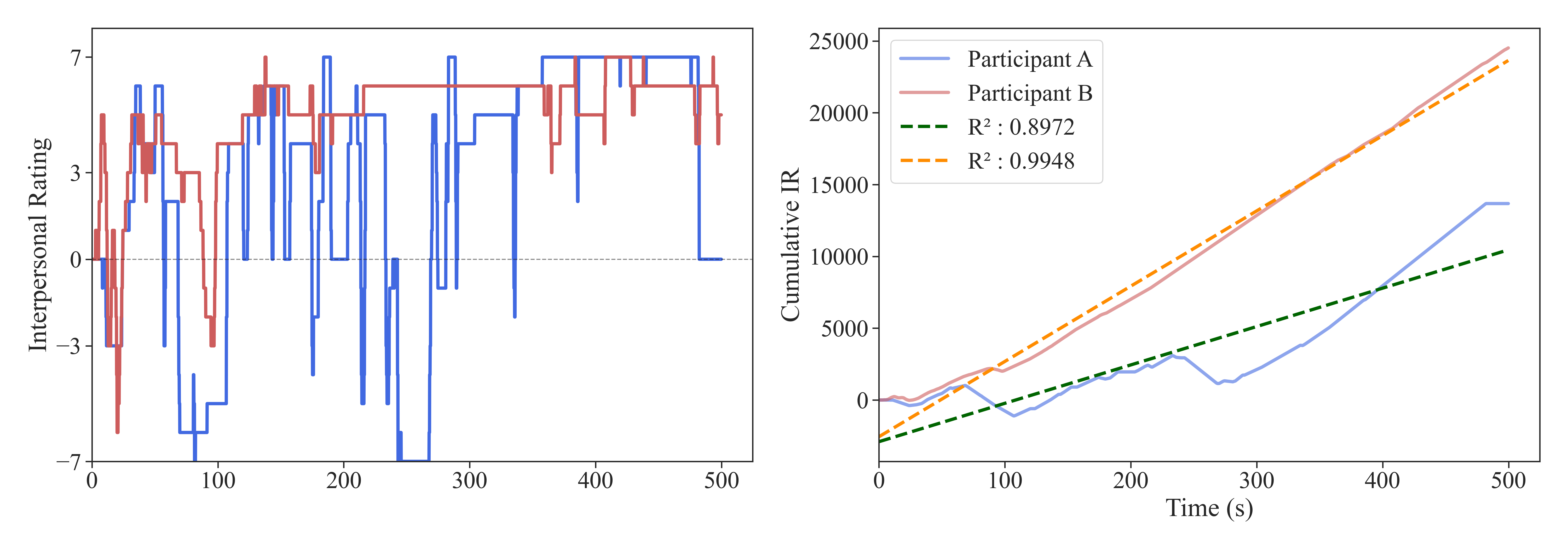}
    \caption{Dynamics of interpersonal ratings (IR) in one session (left) and respective cumulative sum of rated values (right). Interpersonal Perception is calculated by adjusting a linear regression to the latter curves, and extracting the slope of these lines (right, overlaid dashed curve).}
    \label{fig:continuousrate}
\end{figure*}

\begin{figure*} [h]
    \centering
    \includegraphics[width=0.9\textwidth]{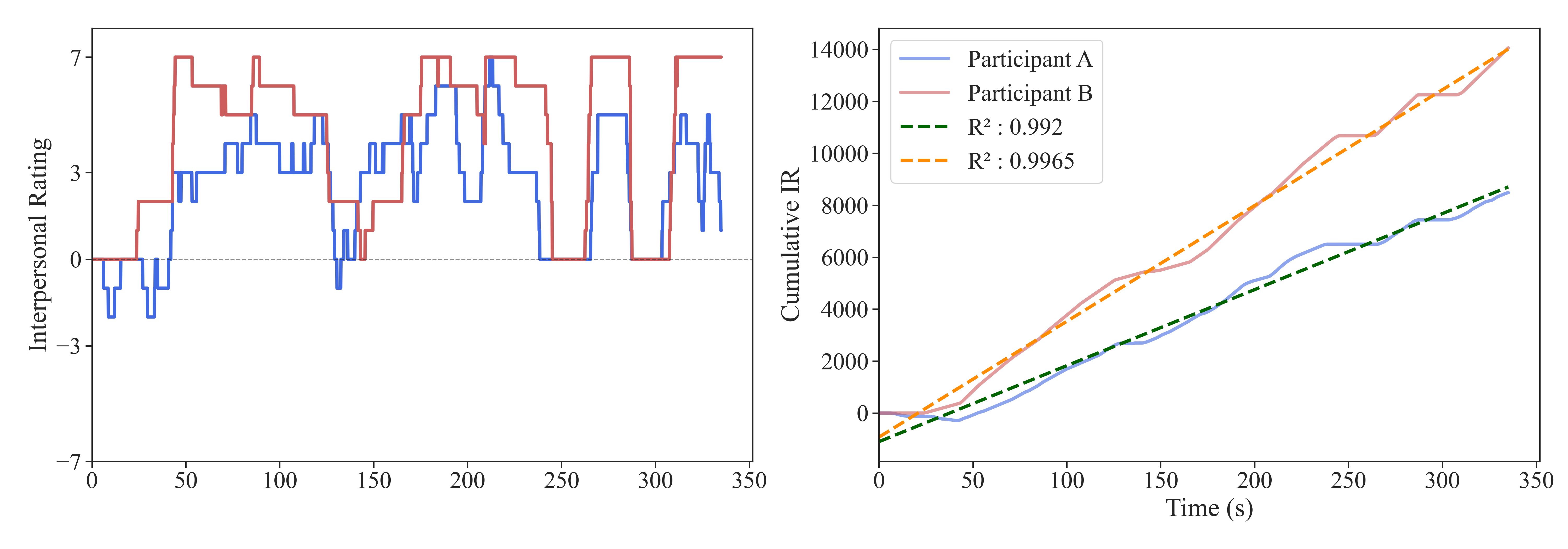}
    \caption{Dynamics of interpersonal ratings (IR) in one session (left) and respective cumulative sum of rated values (right). Interpersonal Perception is calculated by adjusting a linear regression to the latter curves, and extracting the slope of these lines (right, overlaid dashed curve).}
    \label{fig:continuousrate}
\end{figure*}